\documentclass[aps]{revtex4}
\usepackage{amsmath}
\usepackage{amsfonts}
\usepackage{amssymb}
\usepackage{graphicx}

\begin{document}

\date{26/01/2014}
\title{Inhomogeneous and nonstationary Hall states of the CDW
 \\ with quantized normal carriers.}
\author{S. Brazovskii}
\affiliation{LPTMS-CNRS, UMR8626, Univ. Paris-Sud, Bat. 100, Orsay, F-91405
France} \email{brazov@lptms.u-psud.fr}
\affiliation{International Institute of Physics, 59078-400 Natal, Rio Grande do Norte, Brazil}
\begin{abstract}
 We suggest a theory for a deformable and sliding charge density wave (CDW) in the Hall bar geometry for the quantum limit when the carriers in remnant small pockets are concentrated at lowest Landau levels (LL) forming a fractionally ($\nu<1$) filled quantum Hall state.
 The gigantic polarizability  of the CDW allows for a strong redistribution of electronic densities up to a complete charge segregation when all carriers occupy, with the maximum filling, a fraction $\nu$ of the chain length - thus forming the integer quantum Hall state, while leaving the fraction $(1-\nu)$ of the chain length unoccupied.  The electric field in charged regions easily exceeds the pinning threshold of the CDW, then the depinning propagates into the nominally pinned central region via sharp domain walls.
 Resulting picture is that of compensated collective and normal pulsing counter-currents driven by the Hall voltage. This scenario is illustrated by numerical modeling for nonstationary distributions of the current and the electric field. This picture can interpret experiments in mesa-junctions showing depinning by the Hall voltage and the generation of voltage-controlled high frequency oscillations (Yu.I. Latyshev, P. Monceau, A.A. Sinchenko, et al, presented at ECRYS-2011, unpublished).
 
 \textit{After proceedings of ECRYS 2014}

\textbf{Keywords:}
CDW, QHE, Shapiro steps, NBN, mesa junction, NbSe3
\end{abstract}
\maketitle

\section{Introduction.}
\begin{figure}[h,t,b]
\centering
  \includegraphics[width=0.4\textwidth,height=5cm]{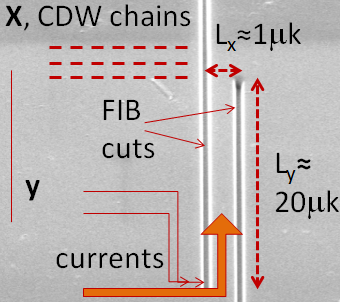}
  \includegraphics[width=0.4\textwidth,height=5cm]{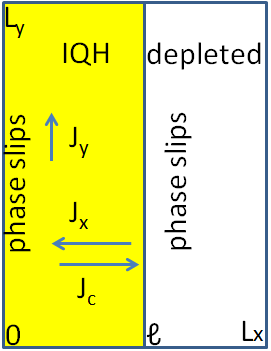}
  \caption{Left panel: The photo \cite{Latyshev-new} of the working simplest mesa structure with comments. Dashed lines show the chains, arrows show the current which is forced to enter the long channel crossing the chains direction.
  Right panel: The scheme of the charge density segregation at T=0 into a region $0<x<l$ of the maximal density in the IQH state and a fully depleted region $l<x<L_x$. Phase slips should take place at the boundary in the sliding regime. $J_y$ is the bias transverse normal current. $J_x$ and $J_c$ are the compensating normal and collective currents in the chains direction.}
  \label{fig:scheme+mesa}
\end{figure}

Commonly, the sliding current of an incommensurate charge density wave (CDW) is driven by the electric field $E_{x}$ in the direction $x$ of chains, which is originated by the applied voltage $V$ and mediated by the normal current density $j_{xn}$. New experiments \cite{Latyshev-new} on mesa junctions in a high magnetic field (HMF) \cite{Latyshev-mesa-HMF-2008,Latyshev-mesa-HMF-2009} show, for the first time, that the CDW can be driven also by the Hall voltage $V_H$ originated by the normal current passing in a transverse direction, Fig.\ref{fig:scheme+mesa}.
The observations are quite counterintuitive with respect to the common notion. In a narrow channel cut transversely to the chains, the conductivity drops below the threshold rather than rising, Shapiro steps are quantized in voltage rather than in current, etc.

Facing these challenges, we shall discuss here the Hall effect in the deformable CDW at the quantum limit of normal carriers. Our other studies show that the regime of classically strong magnetic fields (before installing the quantum Hall regime at lowest LLs) cannot reproduce the complex of experimental dependencies on B and sizes and the scale of the universal Josephson ratio for Shapiro steps. As a minimal model of the basic experimental geometry shown in Fig.\ref{fig:scheme+mesa} (left panel), we shall consider a Hall bar with the current and the magnetic field applied orthogonally in the interchain directions $y,z$.

Small pockets of electrons can be present in the otherwise gapped CDW state.
In a layered system at HMF the carriers are concentrated at one or a few lowest LLs forming fractionally ($\nu <1$) filled QH state. As always, the Lorentz force pushes electrons to one edge $x=0$ of the bar forming the slightly over/under charged boundary layers which give rise to the Hall voltage which strength $E_H=V_H/L_x$ compensates for the Lorentz force.
But now, the gigantic polarizability  of the CDW reduces the electric field  allowing for so strong redistributions of electronic densities which are unthinkable in usual circumstances.
E.g. at low temperature T<1K, the experimental $V_H\approx 1meV$ forces all carriers to completely fill a fraction $\nu$ of the chain length, thus forming there the integer QH state, while leaving the fraction $(1-\nu)$ of the chain length unoccupied.

\section{Overview and results.}
In a (quasi)-2D system at a HMF the electrons form essentially quantum states. But with transitions among the LLs being frozen, the kinetic energy leaves the game, then the density is determined classically by the electric potential $\Phi(\vec{r})$.

At edges of usual QH setups, the density distribution follows almost slavishly the strong Coulomb potential which has been build up already before the HMF was applied; it comes from pronounced finite size effects for the 2D electron gas embedded to the 3D geometry of the device, see \cite{McDonalds,Shklovskii,Thouless}.
But for electrons in pockets upon the CDW, deformations of the CDW phase are able to neutralize any charge so that electrons are allowed to form themselves highly inhomogeneous distributions. Here they are originated intrinsically, to provide the electric field opposing the Lorentz force. At low $T$, that extents to separation of the mean carriers' density
$\bar n_{1}$ among a fully depleted - $n=0$ region where
$\Phi>\Phi_{F}$ and an overfilled one - $n=\bar{n}/\nu$ where $\Phi<\Phi_{F}$,
i.e. between the two, both nonconducting, regions where the same LL is either full or empty.

We consider a stack of plain rectangles of the size $L_{x}\times L_{y}$, each composed of $N_{y}$ chains with a spacing $d_{y}=N_{y}/L_{y}$. Each
chain segment bears a few $N_{1}=\bar{n}_{1}L_{x}$ electrons. In the HMF they are supposed to occupy the lowest LL which can acquire as much as $N_{LL}$ electrons. The mean LL filling is
$\nu={N_{1}}/{N_{LL}}=2\pi\lambda^{2}\bar{n}_{1}/d_{y}$ where $\lambda=\surd(\hbar c/eB)$ is the magnetic length.

The profile for the density of carriers $n_{1}(x)$ is
$n_{1}=f(\Phi-\Phi_{F}){\bar{n}_{1}}/{\nu}$
where $f(\mathcal{E})$ is the Fermi distribution function at a given T and we have assumed that the LL energy $\hbar\omega_{c}/2$ is included to the definition of the Fermi potential energy $\Phi_{F}$. The level $\Phi_{F}$ is determined at the fixed number of
particles by the condition
\[
\int f(\Phi(x)-\Phi_{F})dx=l=\nu L_{x}
\]
The in-plane current density in the direction $y$ is given by
properties of the Landau eigenfunctions in crossed electric and magnetic
fields:
\[
j_{y}(x)=-\frac{e}{h}\Phi^{\prime}(x)f(\Phi(x)-\Phi_{F})
\]
Integration over $x$ gives the overall $y$ current as a function of only values of the boundary potentials:
\begin{equation}
J_{y}=\frac{e}{h}T\ln\frac{\exp[(\Phi_{F}-\Phi(0))/T]+1}
{\exp[(\Phi_{F}-\Phi(L))/T]+1}
\label{J,V0,V1}
\end{equation}

In the limit $T\rightarrow0$ and if all states are filled ($\nu=1$ which
requires that both $\Phi(0),\Phi_{L}<\Phi_{F}$), then we recover the standard
result of the IQHE $J_{y}=(\Phi(0)-\Phi(L))e/h$.
In the limit $T\rightarrow0$ for the partial filling $\nu<1$, the states are
all filled over a segment $(0,l=\nu L_{x})$ and all are depleted beyond, i.e.
$\Phi(0)<\Phi_{F}$, $\Phi_{L}>\Phi_{F}$, $\Phi(l)=0$; then
$J_{y}=(\Phi(0)-\Phi_{F})e/h$ depends explicitly only on the potential at one boundary. But this is only a part of the total voltage: the electric potential keeps growing also in the depleted region $l<x<L_{x}$. The solution of the elementary electrostatic problem yields the value of $\Phi(L)$ at the depleted boundary. Finally the mean current density reproduces the standard result for the classical Hall conductance $\sim1/B$:
\[
J_{y}=\sigma_{xy}(\Phi(0)-\Phi(L))\ ,~\sigma_{xy}=ec\frac{\bar{n}_{2}}{B}
\]
We have got that, in spite of the IQH filling of the populated area, because the division line between the areas is displaced with changing $B$.

By now, there was no voltage drop in $y$ direction which can be only the
dissipative $V_{yy}=\rho_{yy}J_{y}$ with a negligible
$\rho_{yy}=\sigma_{xx}/(\sigma_{xx}\sigma_{yy}+\sigma_{xy}^{2})
\approx\sigma_{xx}/\sigma_{xy}^{2}$. For that, the longitudinal conductivities $\sigma_{xx}$ , $\sigma_{yy}$ should be small with respect to the transverse one $\sigma_{xy}$. In conventional  QH devices it comes from localization of carriers (as the consequence of absence of their kinetic energy) at irregular potential of dopants. For electrons in NbSe3, with its $1D\rightarrow2D\rightarrow3D$ hierarchy, at least the interchain conductivity $\sigma_{yy}$ is very small which ensures the condition $\sigma_{xx}\sigma_{yy}\ll\sigma_{xy}^{2}$ for domination of the
transverse ($\bot\vec{B}$) currents $\vec{j}$ and of the Lorentz force.

 Allowing the CDW to start sliding brings about the normal counter-current $J_{x}=-J_{c}$. The Lorentz force component appears in $y$ direction giving rise to the additional voltage $V_{y}$. Its magnitude is given by the universal law similar to (\ref{J,V0,V1}) with interchanged indices $x$ and $y$ (and the opposite sign). For the current per chain we get (the second line in (\ref{Delta-Vy}) is written for $T=0$):
\begin{eqnarray}
j_{x1}=
\frac{e}{h}\frac{d_{y}}{L_{y}}
\int_{0}^{L_{y}}\partial_{y}\Phi(x,y)f(\mathcal{E}(x,y))dy
\nonumber \\
\approx\frac{e}{h}\frac{d_{y}}{L_{y}}(\Phi(x,0)-\Phi(x,L_{y}))
=\frac{e^{2}}{h}\Delta V_{y}(x)
\label{Delta-Vy}
\end{eqnarray}
Here $\Delta V_{y}(x)$ is the mean interchain voltage drop; literarily, it is concentrated only over the filled stripe $0<x<l$ and it is zero elsewhere while actually the source and the drain edges are expected to be equipotential. Following the experience \cite{Ruzin} from the traditional IQHE, that can be corrected by a consistent $2D$ solution for the distribution of the carries density and the electric field.
By now, we have imposed the perturbation from the $x$ currents upon the already solved case of the bias $y$ current.

The last step is to remind that the CDW countercurrent, per chain, is
$j_{c}=-e\partial_{t}\varphi/\pi$ and it requires for phase slips to annihilate with the normal current $j_{x1}$ at $x=0,l$. Let each phase slip absorbs/releases  $M$
electrons per chain, then the repetition frequency is
\[
f=
\frac{|\dot{\varphi}|}{2\pi M}=
\frac{|j_{x1}|}{2Me}\approx\frac{e|V_{y}|}{2Mh}\ ,\ \frac{|V_{y}|}{f}=\frac{2Mh}{e}
\]
We arrive at the Josephson relation with an additional factor $M$; according
to experiments, with a good accuracy $M\approx10$ which provocatively coincides with the number of electrons per chain. It can happen that each phase slip swallows all electrons at the enriched end to deliver them at the depleted end. Such a strong event cannot be localized only near the ends and the whole dynamical problem needs to be modeled.

\section{NbSe3 in HMF.}
Recall some relevant properties of the CDW in NbSe3. The most important feature is the gigantic dielectric constant $\epsilon$ which source is the CDW polarization. There are several sources in the literature: $\epsilon =5\times 10^{9}$ at T=50K \cite{Thorne}; $\epsilon =2\times 10^{8}$ at T=42K, and $\epsilon =1\times 10^{8}$ at T=20K \cite{Gruner+Zettle}; $\epsilon =5\times 10^{8}$ at T=22K \cite{Seeger}.
We can extrapolate that to lower $T\sim 1K$ with a common wisdom (while not
verified at low $T$) that $\epsilon \sim 1/E_{t}$ which itself is known to be $\sim\exp(T/T_{0})$ with $T_{0}\approx $15.2K \cite{Latyshev-NbSe3-HMF,Coleman}. Taking this law, or just points from plots in \cite{Coleman}, we expect a 20 times reduction from $42K$ to $1K$ and 3 times from $22K$; then we get the low $T$ saturation at
$\epsilon\sim 10^{7-8}$. Notice that $\epsilon\sim 10^{8}$ corresponds to the pinning length  $L_p\approx 1\mu k$, then there will be no bulk pinning at all for our length scales.

 Now we summarize the facts related to normal carriers in NbSe3 at HMF. Hereafter all numerical estimations will be done for $B=20T$.

1. Zeeman splitting is $\Delta_{Z}\approx 2.32meV\approx 25K>V_H$, but still small to fully polarize the pocket of electrons which partial Fermi energy is $6meV$, then $2/3$ of electrons are available to form spinless pairs for an exchange with the CDW condensate.

2. Magnetic length  $\lambda\approx 5.64nm$ is very short in comparison with the chain length $L_x=1\mu m$.

3. There are two types $\alpha$ of carriers: extrinsic $\alpha=e$ which are
electrons and intrinsic $\alpha=i$ which are holes, with effective masses $m_{\alpha}$ in the $x,y$ plane.
$\omega_{c\alpha}=\hbar/(m_{\alpha}\lambda^{2})$ are the cyclotron frequencies
where $\hbar\omega_{ce}\approx11.5meV\approx 12K$ for electrons and an expected bigger value for lighter holes which are for sure in the truly quantum regime, since B>1T \cite{Sinchenko}. We shall suppose that bandwidths in the heaviest $z$ direction are smaller than $\hbar\omega_{c}$ which statement is marginal for electrons and may be ensured for holes. Then electrons or at least holes form a 2D state within every $x,y$ plane.

4. The filling factor $\nu$ of one LL is $\nu=7.8$. Then only one-two LLs are necessary to accommodate the present $\approx 10$ electrons per chain segment and only a small fraction of one LL is sufficient for expected $N_{h}\sim 1$ holes. Precisely for these numbers: for electrons, the $n=0$ LL will be full and the $n=1$ LL will be filled with a fraction $\nu_{e}\approx0.3$; for holes, the $n=0$ LL will be filled with the fraction $\nu_{h}\approx 1/8$.

\section{The modeling.}
\subsection{Conjections.}
We shall proceed with a minimalistic model which concentrates upon only one type of carriers at a zero LL; the  spin splitting is not taken into account.  We shall ignore the special low T states: the fractional QHE at special rational values of $\nu $ and  the Wigner crystallization (its source, the long range Coulomb repulsion, is strongly suppressed in presence of the CDW). Electronic interaction will be taken into account only via the self-consistent slowly varying Coulomb potential $\Phi $. This Hartry approximation was used effectively \cite{McDonalds,Shklovskii,Thouless} in theory of inhomogeneous IQHE near gates boundaries in semiconductors.

 We shall use the parameters as: chain length $L_{x}=1\mu k$, pinning length $l_{p}=0.1\mu k$ (corresponding to $\epsilon\sim 10^{6-7}$), temperature $T=0,0.1,0.4 \ meV$, magnetic field $B=20T$, number of particles $N_{1}=1$ per $L_{x}$ (like for holes), filling factor of the LL as $\nu =0.1$ and $\nu =0.3$. We tested $V_{H}=0.5meV$ and $V_{H}=3meV$ corresponding to the threshold voltage and to the maximal one in experiments.

\subsection{Static regime of pinned CDW.}

For static distributions, the analytic treatment is possible at T=0 which may be applicable at the lower experimental limit $T=1K$. But for higher  $T>V_H$, we should use numerical solutions.
The $T=0$ solution confirms the full charge segregation which sets in when
\[
V_{H}>V_{H}^{cr}=2\pi(1-\nu)\frac{e^{2}}{\epsilon d_{z}}
\frac{L_{x}}{d_{y}}N_{1}
\]
The critical value $V_{H}^{cr}\approx 1meV$ at $\epsilon=10^{7}$ is the energy of Coulomb interaction between the over- and under charged areas.

\begin{figure}[htb]
\includegraphics[height=5cm,width=0.4\textwidth]{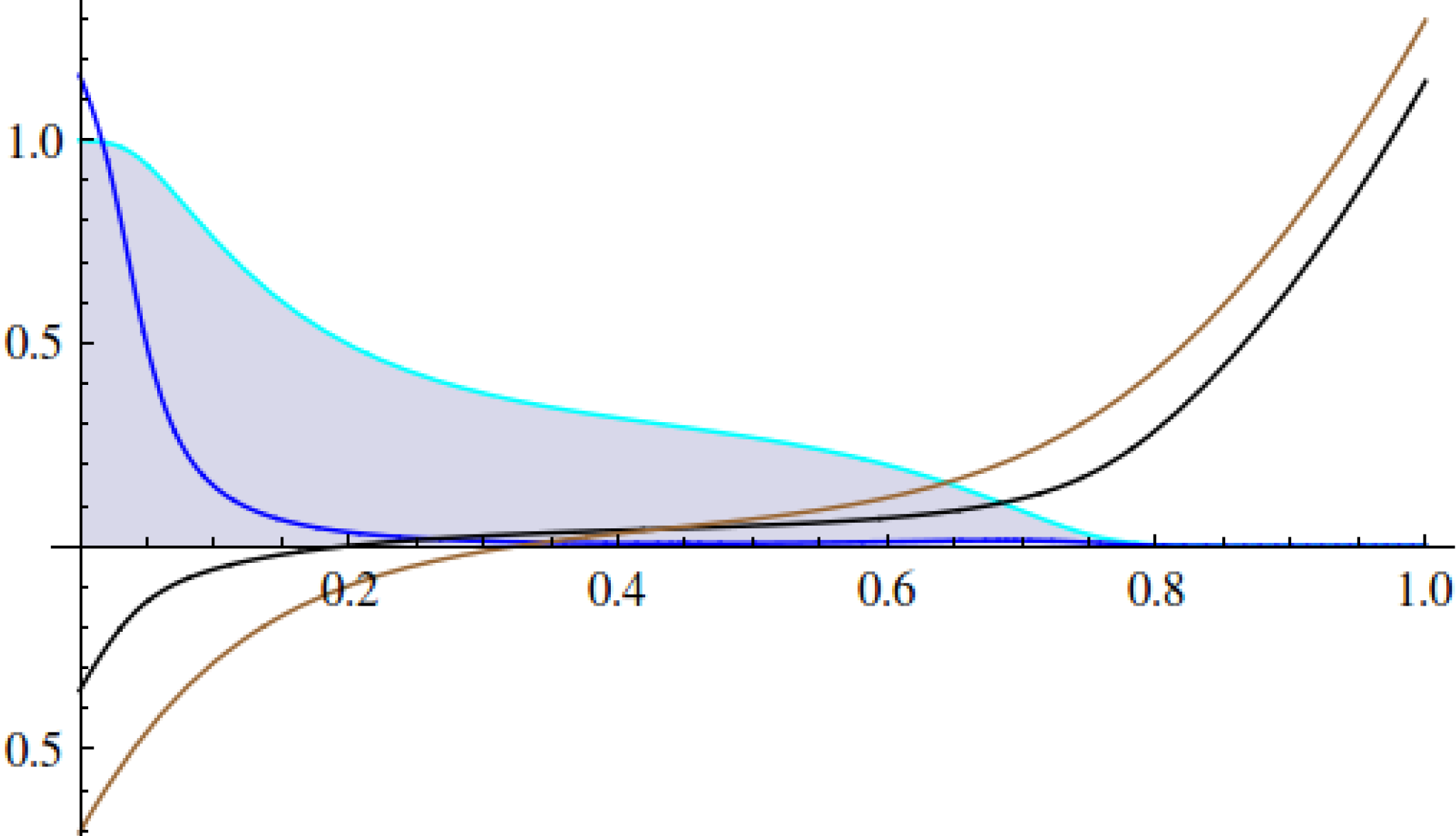}
{\includegraphics[height=4cm,width=0.4\textwidth]{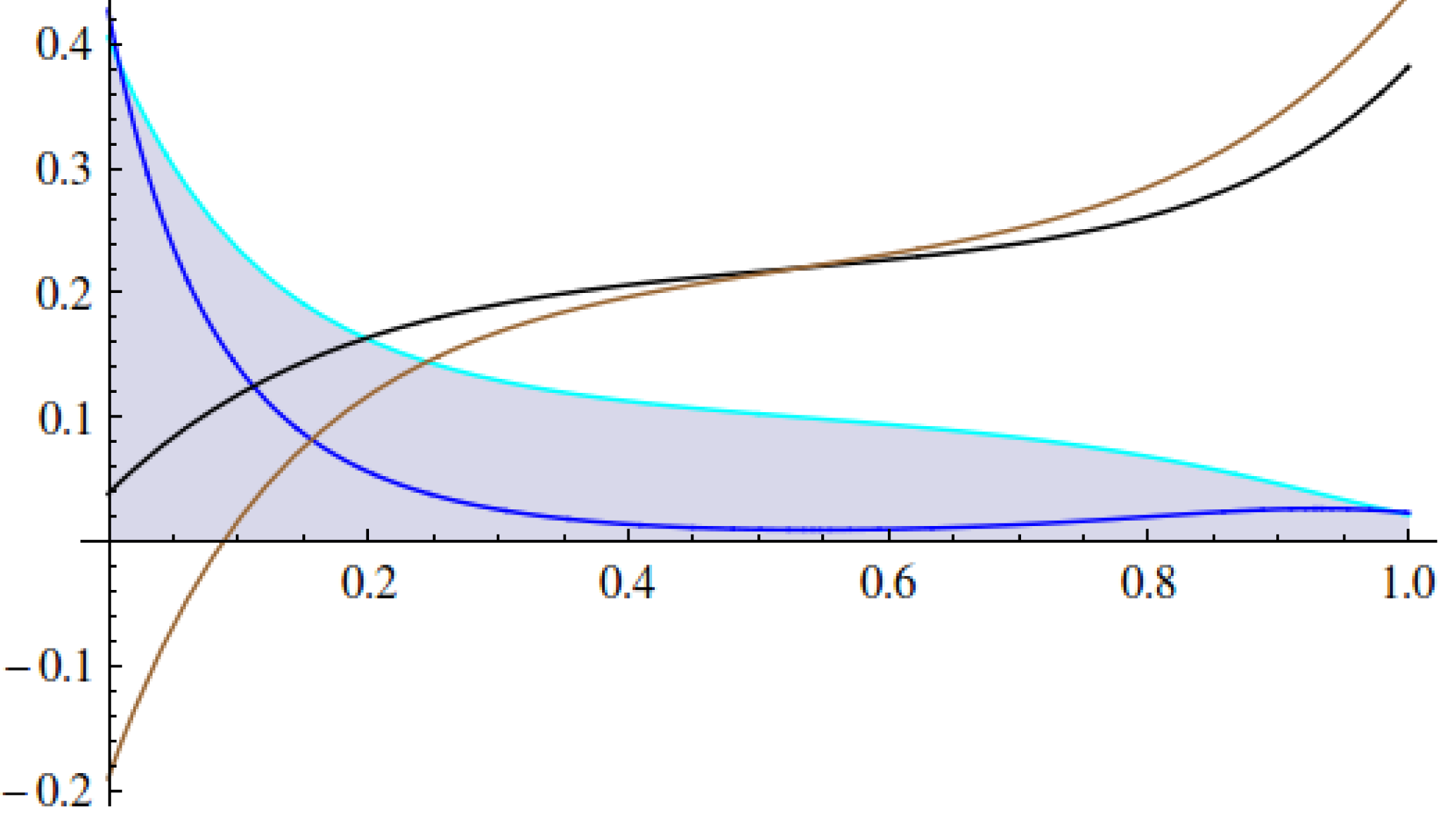}}
\caption{Static profiles for occupation numbers (cyan), currents (blue), electric potentials (black) and the CDW stress $\mu_c$. The parameters are $V_H=3$, $L_p=0.1$; left panel: $T=0.1$, $\nu=0.3$; right panel: $T=0.4$, $\nu=0.1$}
\label{fig:x-plots}
\end{figure}

Plots in Figs.\ref{fig:x-plots} show the numerically calculated distributions.

\noindent $\bullet$ Cyan lines with fillings give distributions of
occupation numbers - the Fermi function $f((\Phi -\Phi_{F})/T)$. At $V_{H}=3mV$ the full depletion appears at the high-voltage end $0.8<x<1$ when $T=1.1K$ is low (cf. at $T=0$ the depletion $\nu=0$ would spread over $0.3<x<1$); the IQH limit $f=1$ is reached only near $x=0$. At the elevated $T=4.4K$ and smaller $\nu=0.1$ the population is spread over the whole length, while it is strongly suppressed from $f=0.4$ at $x=0$ to $f=0.02$ at $x=1$.

\noindent $\bullet$  Blue lines give distributions of $j_y$. They follow the rise of both $\Phi^{\prime}$ and $f$ towards $x=0$. But the current can rise also towards depleted edge $x=1$ where the high field can overcome the drop of the density.

\noindent $\bullet$  Black lines give the monotonic potential $\Phi$, with $eE=-\Phi^{\prime}$ rising near both ends and suppressed, together with the current, in the middle.

\noindent $\bullet$  The brown line gives the CDW stress which is also the
electrochemical potential $\mu_{c}$ of the condensate.
 The shape of $\mu_{c}$ is similar to that of $\Phi $ but with a discrepancy
 $\Delta\mu =\Phi -\mu_{c}$ which changes sign near the middle. That means the two reservoirs of electrons are not stable with respect to exchange of particles among them, provided the phase slips are allowed \cite{X-ray-exp,X-ray-theo}. Then the CDW will give up its periods near the depletion region where $\Delta\mu <0$ and will absorb pairs near the overpopulated end where $\Delta\mu >0$.
The discrepancy $\Delta\mu $ is largest at the left end as we see it
clearly at the plot for $\nu =0.1$, $V_{H}=3$ where $\Delta\mu$ reaches $\approx 0.25meV$ compatible with expectations for the phase slip voltage.

Numerical results were obtained from the following equations. The one for the phase is the balance of the stress gradient and the
pinning force $F_{p}$:
\begin{equation}
\partial_{x}\mu_{c}=F_{p}\ ,\ \mu_{c}=\Phi +\frac{q}{\pi N_{F}}\ ,\
F_{p}=\frac{\varphi}{\pi N_{F}L_{p}^{2}}  \label{lin-pin}
\end{equation}
The potential satisfies the Poisson eq. with the total charge density per chain $n_{tot}=(q/\pi +n-\bar{n})$
\begin{equation}
r_{0}^{2}N_{F}\Phi^{\prime\prime}+n_1\left[f(\Phi /T)/\nu -1\right] +q/\pi=0
 \ , \
 \label{pot} \\
 \nonumber
\end{equation}
where $N_{F}=2/(\pi\hbar v_{F})$, ${r_{0}^{-2}}=8e^{2}/(\hbar v_{F} d_{y}d_{z})$, $r_{0}\lesssim 1nm$. Here $q=\varphi^{\prime}$ is the CDW wave number shift, $-\bar{n}$ is the background compensating charge density.

With so small $r_{0}$, eq. (\ref{pot}) is reduced to the
constraint of the local electroneutrality
\begin{equation}
\pi {n_{1}}\left[ f(\Phi /T)/\nu -1\right] +q=0
\label{en}
\end{equation}
Eqs. (\ref{lin-pin},\ref{en}) were solved numerically with boundary conditions
Here the first condition ensures the total electroneutrality while the
second one fixes the Hall voltage.

\subsection{Time-dependent regime with depinning.}
We extend the above modeling in two respects.

\noindent 1. We shall model the dependence $F_{p}(\varphi)$ in such a way that the linear law (\ref{lin-pin}) sharply drops to zero at $\varphi>\varphi_{p}$
- the  threshold phase displacement  $\varphi_{p}$ of depinning:
\begin{equation}
F_{p}(\varphi)=\frac{\varphi/(\pi N_{F}L_{p}^{2})}
{1+\exp(k(\varphi -\varphi_{p}))}
\end{equation}
This picture corresponds to a conventional first threshold field which looks to be always valid for NbSe3, while a  more complicated two-step depinning is observed at low T in gapful CDWs, see \cite{SB+TN}.

\noindent 2. We shall look for the time dependent evolution starting from the initial moment when the current $J_y$ is turned on. We shall not limit the CDW current, given by the phase velocity $\omega=\partial_{t}\varphi $, to vanish at the boundary.
We shall suppose that the phase slips concentrated in narrow boundary layers \cite{phase-slips} will do this job. In the future, it should be possible to incorporate that to the model.

Now, the eq. (\ref{lin-pin}) is generalized as
\begin{equation}
\partial_{x}\mu_{c}=F_{p}(\varphi)+\gamma \partial_{t}\varphi
\ ,\ \mu_{c}=\Phi + \partial_{x}\varphi /(\pi N_{F})
\label{mu-c}
\end{equation}
Eqs. (\ref{mu-c}), together with (\ref{en}) where $q=\partial_{x}\varphi$, were solved numerically with the paired boundary condition for the electric field and separate conditions for potentials.
$\partial_{x}\Phi(0,t)=\partial_{x}\Phi(L_{x},t)$,
$\Phi(0,t)-\Phi_{F}=V_{0}\tanh (t/\tau)<0$,
$\Phi (L_{x})-\Phi_{F}=V_{1}\tanh (t/\tau)>0$
where the first condition ensures the total electroneutrality while others turn on the boundary voltage values gradually, over a short time $\tau$.
Actually $V_{0}$ and $V_{1}$ are not fixed separately; rather the given current determines the relation (\ref{J,V0,V1}) among them. Notice that for low $T$ when both
$|V_{0}|,V_{1}\gg T$ it fixes only $\Phi (0,t)-\Phi_{F}=J_{y}h/e$ leaving
$\Phi(L_x,t)$ to be evaluated.

Finally, we exclude $\Phi$ from eqs. (\ref{en},\ref{mu-c}) to arrive at the eq. for the phase alone:
\begin{eqnarray}
{\partial_{x}^{2}\varphi}\left(\frac{n_{1}N_{F}T}
{\left(n_{1}-\partial_{x}\varphi/\pi\right)
\left(\partial_{x}\varphi/\pi+n_{1}(1-\nu)\right)}+1\right)
 \nonumber
 \\
-\frac{\varphi/L_{p}^{2}}{(\exp (k(\varphi-\varphi_{p}))+1)}
=\gamma \partial_{t}\varphi
 \nonumber
 \label{chi-xt}
\end{eqnarray}
with $V_{1}-V_{0}=V_{H}$. Remarkably, such a nonlinear nonlocal partial-differential equation yields
reliably a complicated spacio-temporal behavior described below.

The results presented in Fig.\ref{fig:plots(x,t)} were obtained for $L_{p}=0.1$ , $\varphi_{p}=0.1$, $k=1000$, $\Phi_{0}=-1$, $\Phi_{1}=2$.
Also we put $\gamma=1$ which fixes the time scale.

\begin{figure}[htb]
\centering
\includegraphics[height=5.0cm,width=0.4\textwidth]{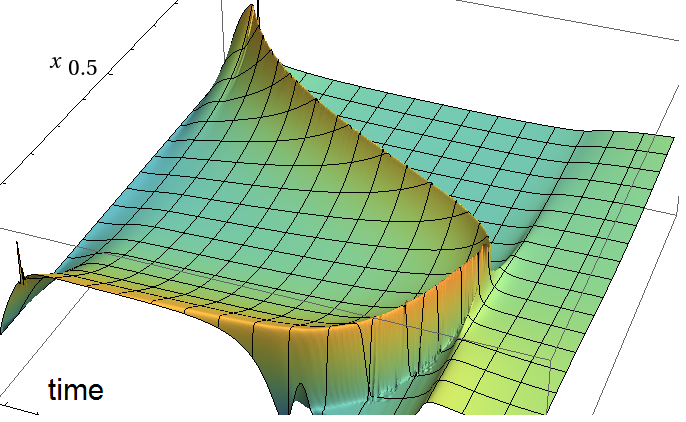}
\includegraphics[height=5.0cm,width=0.4\textwidth]{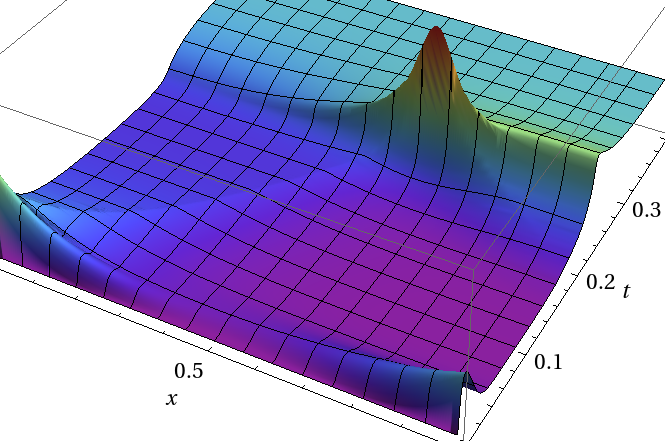}
\caption{Space-time distributions of the electric field (left panel) and the phase velocity (right panel).}
\label{fig:plots(x,t)}
\end{figure}

The left panel of Fig.\ref{fig:plots(x,t)} shows a plot of the electric field $E(x,t)$. It demonstrates two domain walls of the propagating depinning which collide and annihilate giving rise to a stationary sliding depinned state with essential but smooth variations over the length. The plot at the right panel shows that the phase velocity is not as sensitive to the passing walls but gives a strong, very narrow in time peak at their collision.

The interpretation is that electric fields (created by electrons'
redistributions near the boundaries) exceed the threshold field while the
bulk is still below that. The depinning penetrates into the pinned region by
propagating sharp fronts of the electric field. An overshoot will be noticed if we plot also the distribution $n(x,t)$: behind the fronts the occupation numbers are flat at 1 (the transient IQHE state) or zero (the full depletion). But later on the smooth profiles are established.

By now, we studied only the single run (after switching on of the boundary
potentials) assuming a source and a drain for the CDW current by phase slips. The implementation to the Hall bar and particularly to observed oscillations is anticipated to be the following, recall Sec.2. The CDW sliding is arrested at the boundaries by  conversion of particles between the condensate and the normal states which job is made by phase slips. The rate is governed by difference of chemical potentials of the two reservoirs (notice the discrepancy between black and brown lines in plots of the Fig.\ref{fig:x-plots}). We expect that each impulse of the phase slip will reduce  the conditions to the bare ones and the development similar to the above modeling will repeat itself sequentially.

It seems that emergence of depinning waves in our modeling may be of an
significant insight to other, more conventional experimental situations on sliding CDWs.

\section{Outlook and conclusions}
In the bulk of sliding CDWs, the normal $J_{n}$ and the collective $J_{c}$ currents flow in the same direction while regulating their participation ratio depending on the geometry and the background \cite{X-ray-theo}. A curious question may be posed: if the partial currents can flow to opposite directions? Even more, if they can be compensated to a limit when the total current is zero? It happens to be possible indeed in new circumstances when the on-chain
currents are driven by the Hall voltage in a geometry with the closed circuit
in a narrow $x$ direction of the Hall bar, as described in this article and most probably already seen in the experiment \cite{Latyshev-new}. Actually, the old experiments \cite{thermo-sliding} on sliding under the temperature gradient alone should also be understood as the case of compensating currents.

The picture of counter-currents (the collective and the normal ones) closing the loop by annihilating at the Hall bar boundaries looks rather fantastic and even suspicious as a \textit{perpetuum mobile}. Still it is feasible, as well forgotten fundamentals of the conventional Hall effect in semiconductors \cite{Landauer} can tell us.
In this old science there is a silently supposed and well forgotten feature of processes with more than one kind of carriers:
presence of partial Hall currents for electrons and holes; the currents are compensated so that only the total one is zero.
There, the currents of electrons and holes should be compared with the normal and the collective currents of the CDW. In both cases the equilibration among the carriers is hindered: by necessity of the energy relaxation across the gap in semiconductors or of making phase slips in the CDW.

The CDW has a drastic influence already in the stationary regime, before the onset of sliding. Recall the theory \cite{McDonalds,Shklovskii,Thouless} of inhomogeneous IQHE near gates boundaries in semiconductors. In those cases the stripe structure of electronic density appeared in response to the built-in electrostatic field spread from the gate boundary which is present already at $B=0$. In our case, the inhomogeneous distribution of concentration of electrons is caused internally by the Hall voltage. Any spread density profile deviating from the background charge of donors and the gate would be prohibited in a 2D electron gas in semiconductors because of the very high cost of Coulomb energy. In our case, the deformations of the CDW background allows to reduce the field by the gigantic dielectric constant $\epsilon$. The redistribution of charge caused by the Hall voltage alone, up to segregation into the overcharged fully filed IQH segment and the undercharged fully depleted one, is unthinkable in usual circumstance of the QHE.

\medskip

\textbf{Acknowledgements.}
\newline
This article is devoted to memory of Yuri I. Latyshev.
 The author appreciates numerous discussions with P. Monceau and A.A. Sinchenko and useful comments from S.V. Zaitsev-Zotov.

\end{document}